  \documentclass[aps,prl,twocolumn,showpacs]{revtex4}
  \usepackage{graphicx}%
\begin{document}

\title{\bf High Magnetic Field Microwave Conductivity of 2D Electrons in an
Array of Antidots}
\author{P. D. Ye$^{1,2}$, L. W. Engel$^1$, D. C. Tsui$^2$,
J. A. Simmons$^3$, J. R. Wendt$^3$, G. A. Vawter$^3$, and  J. L. Reno$^3$}

\address{$^1$National High Magnetic Field Laboratory, Tallahassee,
Florida 32310}
 
\address{$^2$Department of Electrical Engineering, Princeton University,
Princeton, New Jersey 08544}
 
\address{$^3$Sandia National Laboratories, Albuquerque, New Mexico 87185}

\date{\today}

\begin{abstract}
We measure the  high magnetic field ($B$) microwave conductivity, $\mbox{Re}(\sigma_{xx})$, of a high
mobility 2D electron system  containing an 
antidot array. 
$\mbox{Re}(\sigma_{xx})$ vs  frequency ($f$) increases strongly in the
regime of the  fractional quantum Hall effect series, with Landau 
filling $1/3<\nu<2/3$.  At microwave $f$, $\mbox{Re}(\sigma_{xx})$ vs $B$   
exhibits a broad peak centered around $\nu=1/2$. On the peak, the 10 GHz
$\mbox{Re}(\sigma_{xx})$ can exceed its  dc-limit  value by 
a factor of 5.  This enhanced microwave conductivity is
unobservable for temperature $T\gtrsim 0.5$ K, and grows more pronounced as  
$T$ is decreased.  The effect  may be due to excitations    supported by  the antidot
edges, but  different from the well-known edge magnetoplasmons.  
 
\end{abstract}
\pacs{73.43.Lp, 73.63.Kv, 78.67.Hc}
\maketitle

 Collective excitations of a two dimensional electron system (2DES)
patterned with arrays of
microscopic, artificial structures, have long been of particular interest
since
they can be richer than those of unpatterned 2DES, whose
long-wavelength dipolar excitations are constrained by Kohn's
theorem\cite{kohnref} to show no effect of electron-electron
interaction. One role of the artificial structures is to
couple electromagnetic radiation to the 2DES at enhanced wavevector, $q$,
resulting for example in the observation of 2D plasmons\cite{allentsui}
in far infrared (FIR) optical experiments.  In other cases the
artificial structures support excitations not easily understood
as finite-$q$ plane wave modes of the free 2DES.
An array of antidots (small regions from which electrons are excluded)
applied to a 2DES exhibits
such an excitation\cite{firantidot}, an edge magnetoplasmon
encircling the antidots,  at  much lower frequency than the cyclotron
resonance.   We now report a  microwave-frequency excitation of antidot arrays 
that occurs {\em only} in the low temperature ($T$), high magnetic field 
($B$) regime of the fractional quantum Hall effect (FQHE)\cite{fqheorig}.

Patterning 2DES with antidot or other nanostructure arrays has also
resulted  in striking effects in dc transport. Best known are
the  commensurability oscillations\cite{weissantidot} in low  
$B$, which
result from semiclassical ballistic scattering of electrons off 
the array structures. In an antidot
lattice, such ``geometric resonance'' oscillations
appear as resistance peaks at $B$'s where the
cyclotron orbit of radius $R_c$ encircles an antidot or group of
antidots.  In a development of consequence for the understanding  of
the  FQHE, commensurability
oscillations
have also been observed\cite{kang,smet} in high $B$, near Landau filling
$\nu=1/2$.
These  high-$B$ oscillations are naturally described by composite
fermions (CF's)\cite{hcfreview,cfreview},
exotic particles that contain the electron-electron interaction
responsible for the FQHE, and
that can (for $\nu\sim 1/2$) be thought of as an electron bound up
with two flux quanta. Near $\nu=1/2$ the CF's move
in an effective magnetic field, $B_{eff}=B-B_{\nu=1/2}$,
 and the resulting CF cyclotron orbits of radius
 $ R_c=\hbar (4\pi n_{s}^{1/2})/e  B_{eff}$, where $n_{s}$ is the 2DES carrier
density,
can encircle one  antidot or more, resulting in geometric resonance
resistance peaks.

This paper presents high $B$, finite-frequency  measurements of 2DES
patterned with
nanostructure antidot lattices.  The frequency ($f$) is varied
between 0.1 and 10 GHz,  for temperature $T\ge 100$mK, to cover a previously unexplored regime between the
dc and FIR experiments.    For $f$ above about 2 GHz,  over a broad range of $\nu$,
centered around 1/2,  we  find that the measured diagonal conductivity 
of the array, $\mbox{Re}(\sigma_{xx})$, increases with $f$---by 
as much as a {\em factor of five}
at 10 GHz.  Surprisingly, the antidot array in this regime 
has microwave conductivity {\em increasing} with $f$ while 
{\em decreasing} with $T$. The    
enhanced microwave conductivity (1) is present 
only in the FQHE regime $\nu<1$, and at high $f$ 
appears in   $ \mbox{Re}(\sigma_{xx})$ vs 
$B$ as a striking broad peak roughly symmetric around $\nu=1/2$
and (2) requires  $T<0.5$ K, and increases 
as $T$ is reduced down to 100 mK.  In contrast,
the $ \nu\sim 1/2$ dc-limit conductivity in the same sample is  essentially 
$T$-independent below 1 K.      Since the enhanced  
microwave conductivity  is produced by the  antidots, and has $T$ 
dependence so unlike the dc conductivity,  we 
attribute it to states at the antidot edges which are driven directly 
by the microwave field.  
 
Our samples were prepared from GaAs-AlGaAs
heterojunctions where the 2DES was located approximately 120 nm
underneath the sample surface. With the antidots,
after low $T$ red LED illumination,
  carrier density, $n_{s}$, was $ 1.1 \times 10^{11}\,\rm cm^{-2}$.
We also measured a  reference sample from the
same wafer, but without antidots.
Without antidots, 
this wafer had typical, 0.3 K electron mobility
of around $ \mu\approx 3.6 \times 10^6 \, \rm
cm^2/Vs $.

\begin{figure}[t]
      \includegraphics[scale=1]{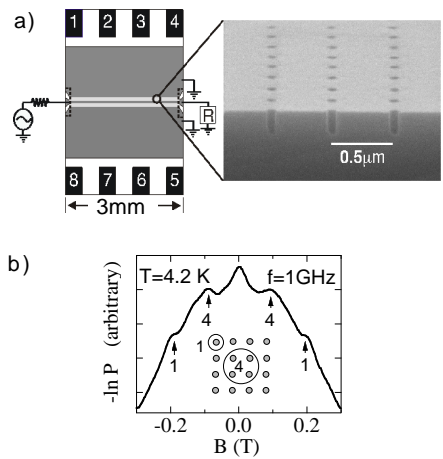}
      \vspace*{-.5in}
\caption{(a) Top view of sample showing
coplanar waveguide transmission line (metal film is dark grey) deposited on top of  
sample.  Ohmic contacts are at the edges (black), and
two antidot
patterned areas (light grey) are in slots. The right part is electron
micrograph of the antidot array.
(b) microwave absorption  for $a=500$~nm antidot lattice.   
Geometric resonances due to  orbits encircling 1 and 4 antidots are 
marked.}
\end{figure}

Fig.~1(a) shows a sketch of the sample with antidots.
On its top surface, 200 \AA\   Ti and 3500 \AA\  
Au were patterned to form a 2 mm long planar   transmission
line, consisting of a 30 $\mu$m wide center strip separated from
side planes by   20 $\mu$m-wide slots.  These slots contain the antidot 
array, a square lattice  with period $a=$500 nm. 
Holes with diameter  50 nm were defined  by electron beam lithography  
and reactive-ion-beam etching\cite{vawter}. 
We report on a sample etched to 50 nm depth;
30 and 70 nm depths gave comparable results.   The depleted region
around an antidot is much larger than the lithographically defined hole in
the wafer;  we estimate the  depletion region  diameter   to be   150
to 250 nm.  The 2DES in the areas enclosed by the dashed lines in
Fig.1(a) was removed  by  wet etching.

The
microwave measurement  methods used here are similar to those described in
earlier
publications~\cite{engel}. $\mbox{Re}(\sigma_{xx})$ is calculated
from the effect of the 2DES  on microwave propagation through the
transmission line which couples capacitively
 to the 2DES.  In the present work, a room-temperature source and receiver
are connected
 to the transmission line by   cryogenic coaxial cables.
 $\mbox{Re}(\sigma_{xx})$, the real part of the
  diagonal conductivity of the 2DES, is related to
the transmitted   power, $P$,
$ \mbox{Re}(\sigma_{xx})=W\vert \ln (P/P_0)\vert /2Z_0d$,
where   $d=2$ mm is
the total length of the transmission line, $W=20 \mu$m is the width
of a slot,   $P_0$ is the transmitted   power for  $\sigma_{xx}=0$, and
$Z_0$ is 50 $\Omega$, which   is the   characteristic
impedance for the  $\sigma_{xx}=0$ case.
 Detailed analysis of the  system in
the quasi-TEM approximation (including reflections, distributed capacitive
coupling, and the effect of the 2DES on the true characteristic
impedance of the line) shows an $f$ (but not $T$ or $B$)
dependent systematic error of about $\pm$ 15 percent
in the $\mbox{Re}(\sigma_{xx})$ taken from this formula.
The determination of $P_0$ produces an additional 
experimental uncertainty in $\mbox{Re}(\sigma_{xx})$ for $f> 2$ GHz.  This
``normalization error'' varies strongly with $f$ but is
independent of $B$ and $T$.  We estimate it as
to  $\pm 0.4~\mu$S for $f>2$ GHz. The in-plane microwave
electric field is largely confined to the slots, so   $\mbox{Re}(\sigma_{xx})$
  characterizes the 2DES in the slots, where the antidot
pattern exists.

At low $B$, both microwave transmission and dc magnetoresistance
show the well-known commensurability
oscillations~\cite{weissantidot}, between   the electron cyclotron radius 
  and the period of the antidot
array. Fig.~1(b) shows 1 GHz
microwave power, $P$, sent through the transmission line, vs $B$
in the low magnetic field range.   Electron geometric resonances
appear for cyclotron orbits
circling one and four 
antidots, indicating that in between the antidots
there is low-disorder 2DES, characterized by mean free path
$\gtrsim 3  \mu$m.

We now focus on  high magnetic fields, especially the range  $2/3<\nu<1/3$, for 
which we obtained our main results. 
Fig.~2(a) shows traces $\mbox{Re}(\sigma_{xx})$ vs  $B$ for
several frequencies $f$. At the lowest $f$ of 0.1~GHz,
the trace shows features   well-known from dc transport,
with several dips in $\mbox{Re}(\sigma_{xx})$ coming from the FQHE,
including the 3/7 and 4/7 FQH states.
Around $\nu=1/2$ ($B=8.5$~T), this trace exhibits a  small peak which is
marked ``CF'' in the figure, since it is likely  an 
effect of  the CF Fermi surface.  This peak is  absent in the   sample
  without the antidots.
We ascribe this small peak  to  the antidots or to
residual disorder induced by the antidot lithography. 
$\mbox{Re}(\sigma_{xx})$ vs $B$ 
shows no  geometric resonances corresponding to  CF cyclotron orbits 
around antidots; observation\cite{kang,smet} of these 
resonances requires lower disorder samples.

\begin{figure}[t]
    \vspace*{-0.1in}
      \includegraphics[scale=0.37]{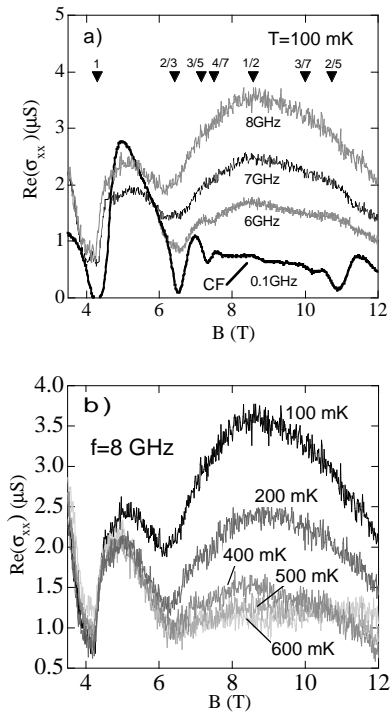} 
\caption{  Measured microwave conductivity, $\mbox{Re}(\sigma_{xx})$ 
vs magnetic field, $B$, for antidot sample. Solid triangles indicate Landau level
filling factors, $\nu$. (a) 100 mk, various frequencies. (b)   
$f=$8 GHz, various temperatures.}
\end{figure}

Our main result appears in  the
successively higher-$f$ traces of Fig.~2(a). Around 8.5 T, where
$\nu=1/2$, $\mbox{Re}(\sigma_{xx})$ vs
$B$ develops a broad maximum, which dominates the conductivity at 
high $f$. 
  $\mbox{Re}(\sigma_{xx})$ clearly
increases monotonically with $f$ in the $B$ range of this maximum
from $B$ just above the 2/3  
FQHE minimum up to our highest $B$ just below the 1/3 effect. 
The $f$ dependence is more complicated
for $B$ just above the $\nu=1 $  integer quantum Hall effect (IQHE),
where  $\mbox{Re}(\sigma_{xx})$
apparently decreases with $f$  between 0.1 and $\sim $ 4 GHz.
The data in Fig. 2(b)  show  that the strong frequency response 
of the conductivity only exists for $T<0.5$ K.
Fig. 2(b) shows   $\mbox{Re}(\sigma_{xx})$ vs $B$ for  $f=8$ GHz,  measured
at several temperatures. The
broad $\nu\sim 1/2$ maximum in $\mbox{Re}(\sigma_{xx})$ vs $B$
disappears gradually as the temperature is
increased; the peak height
is roughly halved at 200 mK, and the peak is unobservable at 600 mK.

In Fig.~2(a), the normalization error
is small for the 0.1 GHz trace, and  can only shift  the other traces
by   $B$-independent constants, estimated to be within   $\pm 0.4~\mu$S.
Normalization error, which is independent of $T$ as well as of $B$
could
uniformly shift all the curves together in Fig. 2(b) by $\pm 0.4~\mu$S.

Fig.~3 summarizes the trends evident from Fig.~2.
 $\mbox{Re}(\sigma_{xx})$ vs $f$ at $T\approx $ 100 mK  is plotted
in Fig.~3(a) for $\nu=1/2, 2/3, 3/5$ and $2/5$.
For the reference sample lacking antidots (open symbols, dotted line), the
error bars shown are
mainly due to normalization error, and  $\mbox{Re}(\sigma_{xx})$ vs $f$
is {\em constant} within this error. 
 The strong $f$-dependence
  is clearly associated with the antidots.  Both 
 the reference sample and the antidot sample have  some $f$ 
 sensitivity at the edges of and around 
 transitions between QHE minima, but only the antidot sample  shows 
 the increasing  $\mbox{Re}(\sigma_{xx})$ vs $f$  in the $1/3<\nu <2/3$ regime. 
The strong $T$ dependence
of the high $f$ conductivity is summarized in
Fig.~3(b). $\mbox{Re}(\sigma_{xx})$ vs $T$ is shown at $\nu=1/2$
($B=8.5$~T ) for $f=$0.1, 4, 8 GHz. 
For $T>0.5 K$, the   conductivity is  nearly the same at microwave 
frequencies and at 0.1 GHz. 
As $T$ is decreased below 0.5 K, the contrast between low-$f$ and 
microwave conductivity is striking: $\mbox{Re}(\sigma_{xx})$   is  nearly
$T$-independent at low
$f$, but at  higher $f$ increases strongly with decreasing
$T$.

\begin{figure}[t]
    \includegraphics[scale=0.35]{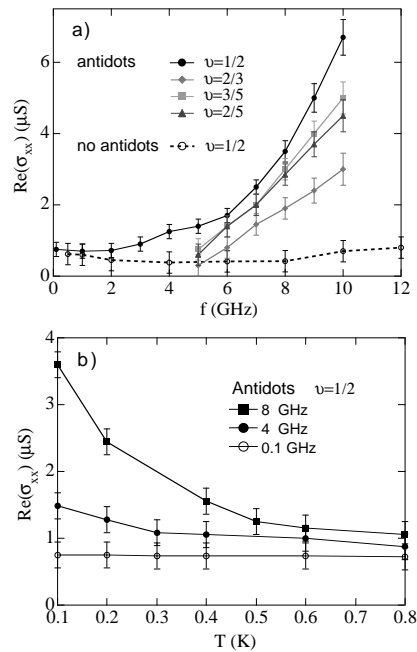} 
    \vspace*{-0.05in}
\caption{ Conductivity $\mbox{Re}(\sigma_{xx})$ for sample with antidots (closed
symbols) and $\nu=1/2$,  and  for sample
without antidots (open symbols). (a)  $\mbox{Re}(\sigma_{xx})$ vs frequency $f$ for 
various Landau fillings $\nu$.  Data for $\nu=2/3,2/5$ and 3/5 below
5 GHz are within error of 0 $\mu$S, and are omitted for clarity. Lines are
guides to the eye.
(b) $\mbox{Re}(\sigma_{xx})$  vs T at $\nu =1/2$, for several frequencies.}
\end{figure} 
 
The $f$ dependent conductivity of CF's or magnetically scattered electrons
may be described with a Drude model, with characteristic scattering
time $\tau$.  At $\nu=1/2$, the observed complex conductivity
$\sigma_{xx}$ is related to the CF conductivity $\tilde{\sigma}_{xx}$
by $\sigma_{xx}={\sigma_{1/2}^{2}}/\tilde{\sigma}_{xx}
$\cite{hcfreview}, where
$\sigma_{1/2}=2e^{2}/h$.   The Drude model
gives $\tilde{\sigma}_{xx} =\tilde{\sigma}_{dc}/(1+i\omega \tau)$, 
where $\omega=2\pi f$,
so observed $\sigma_{xx}=(1+i\omega \tau) \sigma_{dc} $. Thus,
$\mbox{Re}(\sigma_{xx})$ {\em constant} with $f$ is called for by the Drude
model
for CFs, contrary to our experimental results on antidot lattices.
Recent theoretical work~\cite{wilke} applicable to $\nu=1/2$ CF transport
and to magnetic
scattering of electrons in a disorder potential, points out corrections to the
Drude model. For $\omega \tau << 1$, and weak random potentials,
that theory predicts increasing  $\mbox{Re}(\sigma_{xx})$ vs 
$f$ at $\nu=1/2$.

Delocalized CF's  must exist between the antidots to
cause the dc-limit conductivity, but cannot explain the $T$-dependence
 of the high-$f$ peak in $\mbox{Re}(\sigma_{xx})$ vs $B$.
As shown in Fig.~3(b), the low $f$, $\nu=1/2$ conductivity is
essentially $T$ independent for
$T\lesssim 1$ K.   Hence
scattering that leads to dc resistance at $\nu=1/2$
is not   causing
the increase of $\mbox{Re}(\sigma_{xx})$ at high $f$, which 
is strongly reduced even at 200 mK.


Since  the enhanced microwave absorption is produced by the antidots, it
 is reasonable to assume it arises from states  at the antidot edges.  These states   
would couple  to the microwave field, to contribute to  transport at  microwave
frequencies, but not at dc.  Changes in the  configuration of the edge 
states, or in their ability to dissipate microwave power would then be responsible 
for the $T$ dependence of the observed high $f$ $\mbox{Re}(\sigma_{xx})$.  
No peak in the spectrum  is observed, so  the 
measurement can be interpreted as a lossy dielectric response, 
effectively accessing just the low $f$ tail of  edge modes.

The present experiment operates in a different
regime than previous edge 
spectroscopy\cite{firantidot,firdots,rfemp,ashoori,ernstfqh}, and 
 may access edge modes of novel type. The microwave frequency and temperature are
lower than in FIR\cite{firantidot,firdots} investigations, and the 
antidot edges are  much shorter than  the edges
looked at in  rf\cite{rfemp}  or pulsed 
experiments\cite{ashoori,ernstfqh} on
large QHE samples.  Besides  accessing large   wavevectors, the 
smallness of the antidots can make the present measurement sensitive to modes
that would be completely damped out on macroscopic edges.  


The states at the edges of quantum Hall systems
  depend  on the softness of the edge potential profile. 
The edge potential of the etched antidots  in our sample  must
 vary  slowly on the scale of the magnetic length, $(\hbar/eB)^{1/2}$,
and so is   expected to undergo reconstruction into
alternating compressible and incompressible 
strips\cite{chklovskii}. 
Because we observe well-defined FQHE states  
 there are apparently regions between the 
antidots that are well-characterized by bulk states. The antidot 
depletion diameter (150-250 nm) and lattice constant (500nm)
then constrain the length over which strips could develop. 


While much is not 
understood about quantum-Hall edges, tunneling 
experiments\cite{chang,grayson} have 
revealed  behavior of   fractional-regime edges that matches the 
chiral Luttinger liquid 
  description of FQHE edges\cite{wen}, and   
distinct, non-interacting Fermi liquid behavior  of IQHE edges. 
Tunneling experiments measure  I-V curves for tunneling into edges 
of QHE systems; the I-V curves contain a power law regime, in 
which $I\propto V^\alpha$.   Two features of the tunneling results may 
correlate with the present microwave data. First, like the microwave absorption
the tunneling behaves  differently in the  fractional regime, $\nu<1$, 
than it does higher $\nu$.    
With $\nu<1$, $\alpha$ takes the 
value  $(\nu^{-1}-1)$, and  as had been predicted for a  chiral 
Luttinger liquid based
theory\cite{wen}, while $\alpha$ is roughly unity for $\nu>1$ 
\cite{chang}.  
Second,  $\alpha$ has been observed to  vary continuously  according to this formula 
as $\nu$ is swept\cite{grayson}.  This tunneling behavior holds continuously
for  a  range of $\nu$, including 
$1/2$, that is essentially the same   as that for which we observe   
enhanced microwave absorption.

Finally, we cannot rule out more exotic states of the antidot edges as 
explanations of the low $T$ microwave absorption. Edge modes coupled with phonons
in the host semiconductor have
  been predicted theoretically\cite{heinonenphonon}, and would likely 
  be quite sensitive to $T$. 
Edge charge density waves (Wigner
crystal)~\cite{wcrefs} have been predicted to develop  at QHE edges, 
and could be $T$ sensitive.
Coupling
between edges of neighboring  antidots could play a
role in modes of an antidot array  like that studied here.

In conclusion, we observe an anomalous microwave response vs $f$ on
antidot patterned 2DES at high magnetic fields. The observed
$T$-dependence suggests that the effect is associated
with the edges of the antidots.

We thank N. Bonesteel and Kun Yang for valuable
discussions, and Jian Wang and Jie Yao for   assistance.
This work is supported by the Air Force Office of Scientific Research, 
and the National Science Foundation.

\ \\


\begin{thebibliography}{9}

\bibitem{kohnref}W. Kohn,
 Phys. Rev. {\bf 123}, 1242 (1961).

\bibitem{allentsui}
S. J. Allen, Jr.,  D. C. Tsui,  and R. A. Logan,
Phys. Rev. Lett. {\bf 38}, 980 (1977).

\bibitem{firantidot}
K. Kern, {\em et  al.} 
Phys. Rev.
Lett. {\bf 66}, 1618 (1991);
A. Lorke, J. Kotthaus, and K. Ploog, Superlatt. Microstruct. {\bf 9},
103 (1991);
Y. Zhao {\em et  al.},
Appl. Phys. Lett. {\bf 60}, 1510 (1992).
 
\bibitem{fqheorig}D. C. Tsui, H. L. Stormer, and A. C. Gossard,
Phys. Rev. Lett. {\bf 48}, 1559 (1982).
\bibitem{weissantidot}
D. Weiss {\em et al.}, Phys. Rev. Lett. {\bf 66}, 2790 (1991).


\bibitem{kang}W.  Kang {\em et al.},  Phys. Rev. Lett. {\bf 71}, 3850 (1993).
\bibitem{smet} J. H. Smet {\em et al.}, Phys. Rev. B {\bf 56}, 3598 (1997).
\bibitem{hcfreview}B. I. Halperin,
in
 {\em Perspectives in quantum Hall effects}, edited by S. Das Sarma and A.
Pinczuk (Wiley, New York,
1996), 225;  B.I. Halperin,  P.A.  Lee,  and N. Read,
Phys. Rev. B {\bf 47}, 7312 (1993).
\bibitem{cfreview}H. L. Stormer and D. C. Tsui,
in {\em Perspectives in quantum Hall effects},
 edited by S. Das Sarma and A. Pinczuk (Wiley, New York,
1996), 385.


\bibitem{vawter} G.A. Vawter, 
in {\em Handbook of
Advanced Plasma Processing Techniques}, edited by R. Shul and S. Pearton
(Springer, Berlin, 2000), 507.

 \bibitem{engel}
L.W. Engel, D. Shahar, \c{C}. Kurdak, and D.C. Tsui, Phys. Rev. Lett. {\bf 71},
2638 (1993); also  in
{\em Physical Phenomena  in High Magnetic Fields II, proceedings},
(World Scientific, Singapore, 1996) 23.







 


\bibitem{wilke}
J. Wilke {\em et al.},
Phys. Rev. B {\bf 61}, 13774 (2000).
 
\bibitem{firdots} 
 T. Demel {\em et al}, 
 Phys. Rev. Lett. {\bf 64}, 788 (1990).

 \bibitem{rfemp} 
see, for example, I. Grodnensky, D. Heitmann, and K. von Klitzing,
Phys. Rev. Lett. {\bf 67}, 1019 (1991);
N.Q.  Balaban {\em et al.},
Phys. Rev. B { \bf 55}, 13397 (1997).

\bibitem{ashoori}
R. C. Ashoori {\em et al.}, Phys. Rev. B {\bf 45}, 3894 (1992).

\bibitem{ernstfqh}
G. Ernst {\em et al},
Phys. Rev. Lett. {\bf 79}, 3748 (1997).

\bibitem{chklovskii}
 D. B. Chklovskii, B. I. Shklovskii, and L. I. Glazman,
 Phys. Rev. B {\bf 46}, 4026 (1992).

\bibitem{chang}A. M. Chang, L. N. Pfeiffer, and K. W. West, 
Phys. Rev. Lett. {\bf 77}, 2538 (1996).
\bibitem{grayson} M. Grayson {\em et al.},
 Phys. Rev. Lett. {\bf 80}, 1062 (1998).
\bibitem{wen}X. G. Wen,  Phys. Rev. B {\bf 41},12 838 (1990).



\bibitem{heinonenphonon}
O. Heinonen and S. Eggert, Phys. Rev. Lett. {\bf 77}, 358 (1996).

\bibitem{wcrefs}
Eyal Goldmann and Scot R. Renn,
Phys. Rev. B {\bf 60}, 16 611 (1999);   
S. M. Reimann {\em et al.},
Phys. Rev. Lett. {\bf 83}, 3270 (1999). 
 \end{thebibliography}
\end{document}